\documentclass[12pt]{iopart}
\usepackage{amsthm,amssymb,nath}
\usepackage{graphicx}
\let\epsilon\varepsilon
\def\hm{\hphantom{-}}
\def\eqref#1{{\rm(\ref{#1})}}
\newtheorem{proposition}{Proposition}

\newtheorem{remark}{Remark}

\begin{document}

\jl{1}

\title[Integrable Weingarten surfaces]{Classification of  
  integrable Weingarten surfaces possessing an 
  $\mathfrak{sl}$(2)-valued zero curvature representation}
\author{Hynek Baran and Michal Marvan}
\address{Mathematical Institute in Opava, Silesian University in
  Opava, Na Rybn\'\i\v{c}ku 1, 746 01 Opava, Czech Republic.
  {\it E-mail}: Michal.Marvan@math.slu.cz}
\date{}


\begin{abstract}
In this paper we classify Weingarten surfaces integrable in the 
sense of soliton theory. 
The criterion is that the associated Gauss equation possesses an 
$\mathfrak{sl}$(2)-valued zero curvature representation with a 
nonremovable parameter.
Under certain restrictions on the jet order, the answer is given 
by a third order ordinary differential equation to govern the 
functional dependence of the principal curvatures.
Employing the scaling and translation (offsetting) symmetry, we 
give a general solution of the governing equation in terms of 
elliptic integrals. 
We show that the instances when the elliptic integrals degenerate 
to elementary functions were known to nineteenth century geometers. 
Finally, we characterize the associated normal congruences. 
\end{abstract}



\section{Introduction}

Already the classical works of nineteenth-century geometers 
established a major connection between differential geometry and 
the theory of partial differential equations. 
Powerful solution-generating techniques such as the B\"acklund and
Darboux transformations~\cite{R-S} have origins in the prototypical 
relationship between pseudospherical surfaces and solutions of the 
sine-Gordon equation. 

Methods available for solving nonlinear partial differential 
equations were substantially extended in the seventies to include 
the inverse scattering transform and its numerous developments; 
see, e.g.,~\cite{B-B-E-I-M,D-K-N,Ma-S,Z-M-N-P}. 
An important open problem is to describe the class of partial 
differential equations solvable by these powerful methods. 
Indirect detectors such as the symmetry analysis have been 
involved in obtaining extensive complete classifications of 
integrable evolution equations and systems; see~\cite{Mi-S} and
references therein. 
The known theoretical answer given in terms of the existence of 
the associated one-parametric zero curvature representation 
$$
A_y - B_x + [A,B] = 0
$$
has been considered as a classification tool in conjunction 
with the gauge cohomology by one of us~\cite{spp}. 
These methods are not limited to evolution equations, although 
the necessary computations are rather complex, resource consuming, 
and unthinkable without substantial use of computer algebra. 
However, certain partial differential equations of geometric 
origin are particularly well suited for this classification 
method, namely, the Gauss--Mainardi--Codazzi equations of immersed 
surfaces. These equations always possess an associated linear 
zero curvature representation, albeit without the spectral 
parameter. 

Since their introduction by Weingarten~\cite{Wein2}, immersed 
surfaces in $\mathbb R^3$ that satisfy a functional relation 
between the principal curvatures have been of continuing interest 
in differential geometry, see, e.g.,~\cite{H-W,Hopf,K-S}.
It is therefore not surprising that attempts have been made
to identify classes of Weingarten surfaces such that the 
corresponding Gauss equation is integrable in the sense of soliton 
theory. 
The work of Wu~\cite{Wu} and Finkel~\cite{Fin} indicated that all
integrable cases are classical, characterized by a linear relation 
between the Gauss and the mean curvatures (linear Weingarten 
surfaces~\cite[\S812]{DarIII}; 
see also~\cite{G-M-M} and references therein). 
In other words, the integrable Weingarten surfaces were
conjectured to be either minimal or parallel to surfaces of 
constant Gaussian curvature.
This conjecture was, however, disproved by the present authors 
in~\cite{B-M I}, henceforth referred to as Part~I.
In Part I we found another integrable class, consisting of 
surfaces with a constant difference between the principal radii
of curvature, which we called surfaces of constant astigmatism. 
Surprisingly enough, this extra class turned out to be classical
as well, apparently first mentioned by 
Beltrami~\cite[Ch.~9, \S20]{Bel}, covered by Bianchi~\cite{BiaI} 
and Darboux~\cite{DarIII}, see also~\cite{P-Sy}, yet forgotten 
today.

In this paper we continue the work begun in Part~I and complete the 
classification of integrable classes in the simplest possible case.
The integrability criterion we adopt is the existence of an 
$\mathfrak{sl}$(2)-valued zero curvature representation depending 
on a nonremovable parameter.
We apply the same method of formal spectral parameter, introduced 
in~\cite{spp} and briefly reproduced in Part~I.
The underlying symbolic computations, done with the help of 
{\it Maple} and our own package {\it Jets}~\cite{Jets}, are omitted. 
To stay within the limits given by available computing 
resources we had to restrict the jet order (order of derivatives). 

The answer is given by a third order nonlinear ordinary differential 
equation~\eqref{ODE} to govern the functional dependence of the 
principal curvatures. 
Incorporation of the actual spectral parameter is achieved in 
Section~\ref{z:sect}.  
This can be considered a proof of integrability, opening up the 
possibility to obtain explicit solutions by the methods 
of soliton theory~\cite{B-B-E-I-M,D-K-N,Z-M-N-P}.
However, we had to resign ourselves to following this road.
Neither were we able to establish a B\"acklund or Darboux 
transformation~\cite{Ma-S,R-S}, which would allow us to 
construct families of exact solutions depending on an arbitrary 
number of parameters.
We only remark that seed solutions could be conveniently found among 
the rotational surfaces, see~\cite[eq.~(1)]{K-S}.

The governing equation~\eqref{ODE} is explored in 
Section~\ref{sect:ODE}. 
We identify two basic symmetries, scaling and translation 
(offsetting), 
and solve equation~\eqref{ODE} in terms of elliptic integrals.
The generic class of integrable Weingarten surfaces we obtained 
depends on one essential parameter (apart from the scaling and 
offsetting parameters) and is believed to be new.
In Sect.~\ref{sect:summary} we establish the integrable Gauss 
equation~\eqref{Gauss:gen} in the generic case as well as in
a number of special cases when the elliptic integrals degenerate to 
elementary functions. All of these special cases could be located 
in the nineteenth century literature. 

Geometrically, surfaces are related by an offsetting symmetry if 
they are parallel to each other, i.e., if they share the same
normal line congruence.
Therefore, the offsetting symmetry indicates that the concept of 
integrability naturally extends from surfaces to their normal line 
congruences. 
Section~\ref{sect:focal} grew out of our attempt to characterize the 
normal congruences of the integrable Weingarten surfaces.  
We obtain certain relations satisfied by suitably chosen metric 
invariants of the pair of the focal surfaces. 
Naturally, we expect the corresponding focal surfaces to be integrable 
as well, but a detailed investigation had to be postponed to the next 
paper.

\section{Preliminaries}
\label{sect:prelim}

We consider surfaces $\mathbf r(x,y)$, parameterized by the lines of 
curvature. 
This is a regular parameterization except at umbilic points. 
The umbilic points are isolated by the Hartman--Wintner 
theorem~\cite{H-W} except for spheres and planes, which are, therefore, 
the only surfaces excluded from consideration.
 
The fundamental forms can be written as
$$
\numbered\label{ff}
`I = u^2\,`d x^2 + v^2\,`d y^2,  \\
`II = \frac{u^2}\rho\,`d x^2 + \frac{v^2}\sigma\,`d y^2,
$$
where $\rho,\sigma$ are the principal radii of curvature.
The radii transform in a very simple way under the offsetting 
symmetry~\eqref{offset} 
of the integrability problem (unlike the principal curvatures 
$p = 1/\rho$, $q = 1/\sigma$ we used in Part~I). 

Choosing the orthonormal frame
$\Psi = (\mathbf r_x/u, \mathbf r_y/v, \mathbf n)$, we consider the 
Gauss--Weingarten equations
$$
\numbered \label{GW:so3}
\Psi_x = \left(\begin{array}{ccc}
     \hm 0 & -\frac{u_y}{v} & \hm \frac u \rho \\
     \hm \frac{u_y}{v} & \hm 0 & \hm 0 \\
     -\frac u \rho & \hm 0 & \hm 0 \end{array}\right) \Psi,
\qquad
\Psi_y = \left(\begin{array}{ccc}
     \hm 0 & \hm \frac{v_x}{u} & \hm 0 \\
     -\frac{v_x}{u} & \hm 0 & \hm \frac v \sigma \\
     \hm 0 & -\frac v \sigma & \hm 0 \end{array}\right) \Psi
$$
or, more explicitly,
$$
\numbered\label{GW}
\mathbf r_{xx} = \frac{u_x}{u} \mathbf r_x
 - \frac{u u_y}{v^2} \mathbf r_y
 + \frac{u^2}{\rho} \mathbf n, 
\qquad
\mathbf n_{x} = -\frac 1\rho \mathbf r_x, \\
\mathbf r_{xy} = \frac{u_y}{u} \mathbf r_x
 + \frac{v_x}{v} \mathbf r_y, \\
\mathbf r_{yy} = -\frac{v v_x}{u^2} \mathbf r_x
 + \frac{v_y}{v} \mathbf r_y
 + \frac{v^2}{\sigma} \mathbf n, 
\qquad 
\mathbf n_{y} = -\frac 1\sigma \mathbf r_y.
$$
Consequently, the Gauss--Mainardi--Codazzi equations, which are the
compatibility conditions for~\eqref{GW}, read
\begin{equation} \label{GMC-G}
u u_{yy} + v v_{xx} - \frac{v}{u} u_x v_x - \frac{u}{v} u_y v_y
      + \frac{u^2 v^2}{\rho \sigma} = 0,
\end{equation}
and
\begin{equation}
\label{GMC-MC}
\frac{u_y}{u} + \frac{\sigma \rho_y}{\rho (\rho - \sigma)} = 0, \quad
\frac{v_x}{v} + \frac{\rho \sigma_x}{\sigma (\sigma - \rho)} = 0.
\end{equation}

As with Part I, we concentrate on Weingarten surfaces, which are 
characterized by the existence of a functional dependence between $\rho$ 
and $\sigma$.
We often resort to a parametric representation $\rho(w),\sigma(w)$ 
of the dependence. 

Recall that parameters $x,y$ label the lines of curvature; otherwise they 
are arbitrary.  
In line with Finkel's approach~\cite{Fin}, we use this reparameterization 
freedom to solve the Mainardi--Codazzi subsystem~\eqref{GMC-MC}.
The following proposition is a mixture of classical and new results.

\begin{proposition}
\label{prop:uv}
Away from umbilic points, a Weingarten surface can be parameterized by the 
lines of curvature in such a way that
$$
\numbered\label{eq:uv}
u = \exp \int \frac{\rho'\sigma}{(\sigma - \rho)\rho} \,dw, \qquad
v = \exp \int \frac{\rho\sigma'}{(\rho - \sigma)\sigma} \,dw.
$$
The Mainardi--Codazzi subsystem~\eqref{GMC-MC} is then identically satisfied,
while the remaining Gauss equation can be written in the compact form
$$
\numbered\label{W:comp}
R_{yy} + S_{xx} + T = 0,
$$
where $R,S,T$ are appropriate functions of the unknown~$w$.
Moreover, the constraint
$$
\numbered\label{eq:uv1}
(\frac1\rho - \frac1\sigma) u v = 1
$$
can be imposed as an additional condition, and then $T = 1/(\sigma - \rho)$.
\end{proposition}

\begin{proof}
Writing $\rho(w),\sigma(w)$ for some function $w(x,y)$, the general solution 
of the Mainardi--Codazzi subsystem~\eqref{GMC-MC} is
$$
u = u_0(x) \exp \int \frac{\rho'\sigma}{(\sigma - \rho)\rho} \,dw, \qquad
v = v_0(y) \exp \int \frac{\rho\sigma'}{(\rho - \sigma)\sigma} \,dw.
$$
Obviously from formulas~\eqref{ff}, the multipliers $u_0(x)$, $v_0(y)$ can 
be removed by an appropriate relabelling 
$\tilde x = \tilde x(x)$, 
$\tilde y = \tilde y(y)$ 
of the surface's curvature lines. 
With $u_0 = v_0 = 1$, we have
$$
u v = \exp \int (\frac{\rho'\sigma}{(\rho - \sigma)\rho}
 + \frac{\rho\sigma'}{(\sigma - \rho)\sigma})\,dw
 = c \frac{\rho\sigma}{\sigma - \rho}, 
$$
where $c$ is an arbitrary constant multiplier. Setting $c = 1$ by the same 
relabelling argument proves the last relation. 

Having resolved the Mainardi--Codazzi subsystem, we are left with the Gauss 
equation~\eqref{GMC-G} alone.
Multiplied by $1/\rho - 1/\sigma$, equation~\eqref{GMC-G} can be written in 
the compact form~\eqref{W:comp},
where 
$$
\numbered\label{RST}
R = \int \frac{\rho'}{\rho^2} u^2\,dw, \qquad
S = -\int \frac{\sigma'}{\sigma^2} v^2\,dw, \qquad 
T = u^2 v^2 \frac{\sigma - \rho}{\rho^2 \sigma^2}.
$$
Substituting $1/(1/\rho - 1/\sigma)$ for $u v$ finishes the proof.
\end{proof}

\section{The classification result}
\label{z:sect}

Employing the Maple package {\it Jets}~\cite{Jets}, we completed 
the computer-aided cohomological classification outlined in Part~I. 
We have no computer-independent proof of the following result.

\begin{proposition}
The third-order ordinary differential equation
$$
\numbered\label{ODE}
\rho''' = \frac{3}{2 \rho'} \rho''{}^2
  - \frac{\rho' - 1}{\rho - \sigma} \rho''
  + 2 \frac{(\rho' - 1) \rho' (\rho' + 1)}{(\rho - \sigma)^2}.
$$
determines a unique maximal class of Gauss--Mainardi--Codazzi equations 
of Weingarten surfaces whose initial $\mathfrak{sl}(2,\mathbb C)$-valued 
zero curvature representation
$$
\numbered \label{0}
A_0 = \left(\begin{array}{cc}
      \frac{`i u_y}{2v} &
      -\frac{u}{2 \rho} \\
      \frac{u}{2 \rho} &
      -\frac{`i u_y}{2v} \end{array}\right),
\qquad
B_0 = \left(\begin{array}{cc}
      -\frac{`i v_x}{2u} &
      -\frac{`i v}{2 \sigma} \\
      -\frac{`i v}{2 \sigma} &
      \hm \frac{`i v_x}{2u}
\end{array}\right)
$$
admits a second order formal spectral parameter under the condition 
that the normal form of the zero curvature representation can depend on 
derivatives of $u,v,\sigma,\rho$ of no higher than the first order.
\end{proposition}

Here and in what follows we assume that $\rho$ is a function of $\sigma$
and the prime refers to derivatives with respect to~$\sigma$. 
A $k$th order formal parameter $\lambda$ means a power series in terms 
of $\lambda$ up to order $k$.
Part~I should be consulted for the other unexplained notions.

\begin{remark} \rm
(1) The last proposition provides a complete classification of integrable 
Weingarten surfaces under the following assumptions:
The one-parametric zero curvature representation takes values in the Lie
algebra $\mathfrak{sl}(2)$, includes the initial zero curvature 
representation~\eqref{0} as a member, depends analytically on the parameter, 
and its normal form involves derivatives of no higher than the first order.
All these limitations can be overcome, in principle~\cite{M2}, at the cost 
of requiring significantly more computational resources.

(2) We would like to stress that the only part relying on machine 
computations is the completeness of the classification.
All the other proofs in this paper are traditional.
\end{remark}


In the rest of this section we establish integrability 
of the class determined by equation~\eqref{ODE}.
The equation itself will be solved in the next section.

\begin{proposition}
The nonremovable spectral parameter exists for all dependences 
$\rho(\sigma)$ allowed by the governing equation~\eqref{ODE}.
\end{proposition}

\begin{proof}
Inspired by the results of the computer-aided classification, we depart from 
the following ansatz for the parameter-dependent zero curvature 
representation:
$$
A = (\begin{array}{cc}
a_{111} \frac{u_y} v + a_{110} \sigma_x & a_{12} u \\
a_{21} u & -a_{111} \frac{u_y} v - a_{110} \sigma_x
\end{array}),
\\
B = (\begin{array}{cc}
b_{111} \frac{v_x} u + b_{110} \sigma_y & b_{12} v \\
b_{12} v & -b_{111} \frac{v_x} u - b_{110} \sigma_y
\end{array}),
$$
with $a_{111},b_{111},a_{110},b_{110},a_{12},a_{21},b_{12}$ being the
unknown functions of $\sigma$.
The problem is to solve the zero curvature condition 
$D_y A - D_x B  + [A,B] = 0$ for matrix functions $A,B$ of
$u,v,\sigma,\rho$ and their derivatives.
However, the derivatives are not independent quantities, being subject 
to the Gauss--Mainardi--Codazzi equations. The proper way to deal with
this situation is to introduce the manifold determined by the equation 
and its derivatives (a diffiety~\cite{B-V-V}). 
This is fairly easy if the order of derivatives is restricted as it is. 
Initially the derivatives are considered to be independent 
(jet space coordinates).
Considering $\rho$ as a function of $\sigma$ and resolving the 
Mainardi--Codazzi equations~\eqref{GMC-MC}
with respect to $u_y,v_x$, we can express $u_y,v_x$ as functions of
$u,v,\sigma,\sigma_x,\sigma_y$. 
Similarly, the derivatives of the Mainardi--Codazzi 
equations~\eqref{GMC-MC} can be resolved with respect to 
$u_{xy},u_{yy},v_{xx},v_{xy}$, giving $u_{xy},u_{yy},v_{xx},v_{xy}$ 
as functions of $u,u_x,v,v_y,\sigma,\sigma_x,\sigma_y$.
Consequently, the Gauss equation~\eqref{GMC-G} can be written in 
terms of 
$u,u_x,v,v_y,\sigma,\sigma_x,\sigma_y,\sigma_{xx},\sigma_{yy}$,
and then resolved with respect to $\sigma_{yy}$. 
The explicit formulas are somewhat cumbersome, hence omitted. 

With $A,B$ chosen as above, the left-hand side 
$S := D_y A - D_x B  + [A,B]$ of the zero  
curvature condition $S = 0$ is a matrix function of $u,u_x,v,v_y,\sigma, 
\sigma_x,\sigma_y,\sigma_{xx},\sigma_{xy}$.
From $\partial S/\partial\sigma_{xx} = 0$ and 
$\partial S/\partial\sigma_{xy} = 0$ we obtain
$$
b_{111} = -a_{111}, \qquad b_{110} = a_{110}.
$$
From either $\partial^2 S/\partial \sigma_x^2 = 0$ or $\partial^2 S/ 
\partial \sigma_y^2 = 0$ we get $a_{111}' = 0$.
Hence, $a_{111}$ is a constant, which we rename $\lambda$ in  
anticipation of its role as the spectral parameter.

Now, $\partial S/\partial \sigma_x = 0$ if and only if
$$
\numbered\label{b'}
a_{110} = \frac{\lambda \rho}{2 \sigma (\sigma - \rho)} \,
  \frac{a_{12} + a_{21}}{b_{12}}, \qquad
b_{12}' =
  \frac{\rho}{\sigma (\sigma - \rho)} [b_{12} + \lambda (a_{21} - a_{12})],
$$
while $\partial S/\partial \sigma_y = 0$ can be rewritten as
$$
\numbered\label{a'}
a_{12}' = 2 a_{110} a_{12}
  + \frac{\sigma \rho'}{\rho (\rho - \sigma)} (a_{12} + 2 \lambda  
b_{12}),
\\
a_{21}' = -2 a_{110} a_{21}
  + \frac{\sigma \rho'}{\rho (\rho - \sigma)} (a_{21} - 2 \lambda  
b_{12}),
$$
Modulo these relations, vanishing of $S$ is equivalent to
$$
\numbered\label{S}
b_{12} = \frac{\lambda}{\rho\sigma(a_{12} - a_{21})}.
$$

We claim that the governing equation~\eqref{ODE} arises as the condition 
that the system~\eqref{b'}, \eqref{a'}, and~\eqref{S} be compatible for  
arbitrary $\lambda \ne 0$.
To prove this, we denote $P = a_{12} + a_{21}$, $Q = a_{12} - a_{21}$.
With $a_{110}$ and $b_{12}$ taken from formulas~\eqref{b'} and \eqref{S}, 
respectively, equations~\eqref{a'} turn into
$$
\numbered\label{P'Q'}
P' = P \frac{\sigma \rho' - Q^2 \rho^3}{\rho (\rho - \sigma)},
\qquad
Q' = Q \frac{\sigma \rho' - P^2 \rho^3}{\rho (\rho - \sigma)}
  + \frac{4 \lambda^2 \rho'}{\rho^2 (\rho - \sigma)} \, \frac1Q,
$$
and the second equation in~\eqref{b'} into
$$
\numbered\label{PQ}
\rho^4 (Q^2 - P^2) Q^2 + \rho^2 (\rho' - 1) P^2 + 4 \lambda^2 \rho' = 0.
$$
Now the question is whether equations~\eqref{P'Q'} and~\eqref{PQ} are  
compatible.
Modulo eq.~\eqref{P'Q'}, the derivative of~\eqref{PQ} with respect to 
$\sigma$ is
$$
\numbered\label{(PQ)'}
2 \rho^6 (P^2 - Q^2) P^2 Q^2
  + 2 (1 - 3 \rho') \rho^4 P^2 Q^2 - 4 \rho' \lambda^2 P^2
\\\quad
  + (4 \lambda^2 + \rho^2 Q^2)
    [4 \rho' \rho^2 Q^2 + (\rho - \sigma) \rho'' + 2 \rho'{}^2 - 2  
\rho'] = 0.
$$
This is equivalent to
$$
\numbered\label{Q}
[(\rho - \sigma) \rho'' - 2 \rho'{}^2 + 2 (1 + 8 \lambda^2) \rho']
\rho^2 Q^2
  + 4 \lambda^2 [(\rho - \sigma) \rho'' - 2 \rho'{}^2 - 2 \rho'] = 0
$$
modulo~\eqref{PQ}, since~\eqref{Q} is the remainder after division 
of~\eqref{(PQ)'} by~\eqref{PQ} as polynomials in~$P$.
Similarly, dividing~\eqref{PQ} by~\eqref{Q} as polynomials in $Q$, we get
$$
\numbered\label{P}
[(\rho - \sigma) \rho'' - 2 \rho'{}^2 - 2 \rho']
[(\rho - \sigma) \rho'' - 2 \rho'{}^2 + 2 (1 + 8 \lambda^2) \rho']
   \rho^2 P^2
\\\quad
  - 4 (1 + 4 \lambda^2) [(\sigma - \rho)^2 \rho''{}^2
      - 4 \rho'{}^4 + 8 (1 + 8 \lambda^2) \rho'{}^3 - 4 \rho'{}^2]
  = 0.
$$
Differentiating~\eqref{(PQ)'} once more and taking the result 
modulo~\eqref{P'Q'},~\eqref{P} and~\eqref{Q}, we get the governing 
equation~\eqref{ODE} immediately.

Summing up, we obtain a zero curvature representation
$$
A = (\begin{array}{cc}
-\frac{\lambda \sigma \rho'}{\rho (\rho-\sigma)} \frac{u}{v} \sigma_y
  - \frac12 \frac{\rho^2}{\rho-\sigma} P Q \sigma_x
&
\frac12 (P + Q) u
\\
\frac12 (P - Q) u
&
\frac{\lambda \sigma \rho'}{\rho (\rho-\sigma)} \frac{u}{v} \sigma_y
  + \frac12 \frac{\rho^2}{\rho - \sigma} P Q \sigma_x
\end{array}),
\\
B = (\begin{array}{cc}
-\frac{\lambda \rho}{\sigma (\rho-\sigma)} \frac{v}{u} \sigma_x
  - \frac12 \frac{\rho^2}{\rho - \sigma} P Q \sigma_y
&
\frac{\lambda}{\sigma \rho Q} v
\\
\frac{\lambda}{\sigma \rho Q} v &
\frac{\lambda \rho}{\sigma (\rho-\sigma)} \frac{v}{u} \sigma_x
  + \frac12 \frac{\rho^2}{\rho - \sigma} P Q \sigma_y
\end{array}),
$$
where $P$ and $Q$ are the square roots to be determined from 
equations~\eqref{P} and~\eqref{Q}, respectively.
Away from umbilic points (where $\rho = \sigma$), matrices $A,B$ actually 
exist unless $(\rho - \sigma) \rho'' - 2 \rho'{}^2 - 2 \rho' = 0$ when 
$P$ is undefined. 
This excludes exactly spheres and the linear Weingarten surfaces.
The latter surfaces are, however, well known to be integrable, being 
parallel to surfaces of constant curvature (either Gaussian or mean), 
see~\cite{Wu} or \cite[\S1.5.2]{R-S}. 

If $\lambda = `i/2$, then we have $P = 0$ and $Q = 1/r^2$, which 
reproduces the parameterless zero curvature representation~\eqref{0}
we started with. 
\end{proof}

Non-removability of the parameter is ensured by the method~\cite{spp}
(follows from nontriviality of the first gauge cohomology group).

\section{Solution of the governing equation}
\label{sect:ODE}

Apart from the discrete symmetry $\rho \leftrightarrow \sigma$, the  
governing equation~\eqref{ODE} has two obvious continuous symmetries,  
which should be expected in every integrable class of surfaces: the  
{\it scaling symmetry}
$$
\numbered\label{rescale}
\rho \mapsto `e^T \rho, \qquad \sigma \mapsto `e^T \sigma
$$
and the {\it translational symmetry}
$$
\numbered\label{offset}
\rho \mapsto \rho + T, \qquad \sigma \mapsto \sigma + T.
$$
The geometric meaning of the latter symmetry is {\it offsetting\/},  
also known as taking the {\it parallel surface}.
In terms of position vectors, $\mathbf r$ is transformed to
$\mathbf r + T \mathbf n$, where $\mathbf n$ is the unit normal vector
and $T$ is the distance.

With the help of these symmetries we can reduce the order of 
equation~\eqref{ODE} by two.
This can be done by rewriting the equation in terms of the symmetry  
invariants.
Since rescaling applies also to the offset, the translational  
reduction should precede the scaling reduction.
For the two lowest-order translational invariants we choose
$$
\numbered\label{xi eta}
\xi = \rho - \sigma, \qquad \eta = \rho'
$$
(recall that the prime denotes the derivative with respect to $\sigma$).

\paragraph{$1$.}
If $\xi' = 0$ (equivalently, $\rho' = 1$), then $\rho - \sigma = `const 
$, which are the surfaces of constant astigmatism we dealt with in Part~I.

\paragraph{$2$.}
Otherwise, more translational invariants can be computed as  
derivatives of $\eta$ with respect to $\xi$:
$$
\numbered\label{eta_xi}
\eta_\xi = \frac{\eta'}{\xi'}
 = \frac{\rho''}{\rho' - 1}, \quad
\eta_{\xi\xi} = \frac{\rho'''}{(\rho' - 1)^2}
  - \frac{\rho''{}^2}{(\rho' - 1)^3},
$$
etc. In terms of these invariants, the governing equation~\eqref{ODE}  
reduces to the second-order equation
$$
\numbered\label{ODE2}
2 \xi^2 (\eta - 1) \eta \eta_{\xi\xi} - \xi^2 (\eta - 3) \eta_\xi^2 +  
2 \xi (\eta - 1) \eta \eta_\xi - 4 (\eta + 1) \eta^2 = 0.
$$
As expected, this equation is scaling invariant. 
To reduce it with respect to scaling, we proceed as follows.
Besides $\eta$, one more scaling invariant is
$$
\numbered\label{zeta}
\zeta = \xi (\eta - 1) \eta_\xi.
$$
Although dispensable, the factor $\eta - 1$ simplifies the computations 
to follow.

\paragraph{$2.1$.}
If $\eta' = 0$, i.e., $\rho'' = 0$, then~\eqref{ODE} reduces to $\rho'  
= c$, where $c$ is either of $-1,0,1$. The corresponding surfaces are,  
respectively, the constant mean curvature surfaces (a~subclass of  
linear Weingarten surfaces), the tubular surfaces (surfaces swept by  
spheres of constant radius moving along a space curve) and once more  
the constant astigmatism surfaces.


\paragraph{$2.2$.}
Otherwise $\rho'' \ne 0$ and we have
$$
\zeta_\eta = \frac{\rho'''}{\rho''} (\rho - \sigma) + \rho' - 1.
$$
In terms of $\eta,\zeta$, the reduced governing equation~\eqref{ODE2}  
becomes the Bernoulli equation
$$
\zeta_\eta = \frac32 \frac\zeta\eta + 2\frac{\eta^3 - \eta}\zeta
$$
with the general solution
$\zeta^2 = 4 (\eta^2 + 2 c_0 \eta + 1) \eta^2$, where $c_0$ is the  
integration constant.
Substituting from eq.~\eqref{zeta} yields the separable first-order  
equation
$$
\numbered\label{eta:xi}
\xi \frac{d\eta}{d\xi} = \pm 2 \frac\eta{\eta - 1} \sqrt{\eta^2 + 2  
c_0 \eta
+ 1}
$$
containing the parameter $c_0$.
Being written in terms of the scaling and translation invariants, 
this equation determines the integrable Weingarten surfaces up to 
rescaling and offsetting.
Depending on the value of the parameter $c_0$ and on the choice  
of the `$\pm$' sign, we obtain the following cases.


\paragraph{$2.2.1$}
Let $c_0 = 1$. Equation~\eqref{eta:xi} becomes
$$
\numbered\label{eta:xi 1}
\xi \frac{d\eta}{d\xi} = \pm 2 \frac{(\eta + 1) \eta}{\eta - 1}.
$$

\paragraph{$2.2.1.1$.}
With the choice of the plus sign in~\eqref{eta:xi 1}, the general solution is
$(\eta + 1)^2 = c_1 \eta \xi^2$.
Substituting from eq.~\eqref{xi eta}, we obtain
$$
(\rho' + 1)^2 = c_1 (\rho - \sigma)^2 \rho'.
$$
If $c_1 = 0$, the general solution is $\rho + \sigma = `const$.
Otherwise, we apply the transformation
$$
\numbered\label{ss}
\kappa = \rho + \sigma, \quad
\xi = \rho - \sigma
$$
to get
$$
(c_1 \xi^2 - 4) (\frac{d\kappa}{d\xi})^2
  = c_1 \xi^2.
$$
The equation is separable with a general solution
$(\kappa - c_2)^2 - \xi^2 + 4/c_1 = 0$, i.e.,
$$
4 \rho\sigma - 2 c_2 (\rho + \sigma) + \frac{4 + c_1 c_2^2}{c_1} = 0.
$$
In both cases, $c_1 = 0$ and $c_1 \ne 0$, solutions correspond to the  
linear Weingarten surfaces.

\paragraph{$2.2.1.2$.}
With the choice of the minus sign in~\eqref{eta:xi 1}, the general solution is
$(\eta + 1)^2 \xi^2 = c_1 \eta$.
Substituting from eq.~\eqref{xi eta}, we obtain
$(\rho' + 1)^2 (\rho - \sigma)^2 = c_1 \rho'$.
For $c_1 = 0$ we have the special linear Weingarten surfaces $\rho +  
\sigma = `const$ again.
Otherwise, we apply the transformation \eqref{ss} to get
$$
(4 \xi^2 - c_1) (\frac{d\kappa}{d\xi})^2 + c_1 = 0.
$$
The solutions are
$$
\kappa = \pm \frac12 \sqrt{-c_1}
\ln(2 \sqrt{-c_1} \xi + \sqrt{c_1^2 - 4 c_1 \xi^2}) + c_2,
$$
where $c_2$ is the integration constant.

\paragraph{$2.2.1.2.1$.}
For $c_1 < 0$ we can write
$$
\xi = \frac{\sqrt{-c_1}}{2} \sinh(\pm\frac 2{\sqrt{-c_1}} (\kappa - c_2)
  - \ln(-c_1))
$$
or
$$
\frac{\rho - \sigma}{C_1} = \pm \sinh(\frac{\rho + \sigma}{C_1} + C_0).
$$

\paragraph{$2.2.1.2.2$.}
Similarly, solutions corresponding to positive $c_1$ are
$$
\numbered\label{SIN}
\frac{\rho - \sigma}{C_1} = \sin(\frac{\rho + \sigma}{C_1} + C_0).
$$

\paragraph{$2.2.2$}
Let $c = -1$. Equation~\eqref{eta:xi} becomes
$$
\numbered\label{eta:xi -1}
(\eta - 1)^2 (\xi \frac{d\eta}{d\xi} - 2\eta) (\xi \frac{d\eta}{d\xi}  
+ 2\eta) = 0.
$$
Solutions corresponding to $\eta = 1$ belong to Case~1 (constant  
astigmatism surfaces).

\paragraph{$2.2.2.1$.}
The general solution of $\xi \frac{d\eta}{d\xi} = 2\eta$ is
$\eta = c_1 \xi^2$.
Substituting from eq.~\eqref{xi eta}, we obtain the Riccati equation 
$\rho' = c_1 (\rho - \sigma)^2$.

\paragraph{$2.2.2.1.1$.}
For $c_1 > 0$ we get
$$
\numbered\label{tanh}
\rho = \sigma - \frac{\tanh(\sqrt{c_1}\sigma + c_2)}{\sqrt{c_1}}
\quad\text{or}\quad
\rho = \sigma - \frac{\coth(\sqrt{c_1}\sigma + c_2)}{\sqrt{c_1}}
$$
according to whether the integration constant is positive or negative.

\paragraph{$2.2.2.1.2$.}
Similarly, for $c_1 < 0$ we get
$$
\numbered\label{tan}
\rho = \sigma - \frac{\tan(\sqrt{-c_1}\sigma + c_2)}{\sqrt{-c_1}}
\quad\text{or}\quad
\rho = \sigma + \frac{\cot(\sqrt{-c_1}\sigma + c_2)}{\sqrt{-c_1}}.
$$

\paragraph{$2.2.2.2$.}
When solving $\xi \frac{d\eta}{d\xi} = -2\eta$, we get~\eqref{tanh}  
and~\eqref{tan} with $\rho,\sigma$ interchanged.

\paragraph{$2.2.3$.}
We are left with the generic case $c_0 \not\in \{-1,1\}$. 
Equation~\eqref{eta:xi} has the general solution
$$
\numbered\label{generic:ode}
(\eta + c_0 + \sqrt{\eta^2 + 2 c_0 \eta + 1})
   (c_0 \eta + 1 + \sqrt{\eta^2 + 2 c_0 \eta + 1})
  = c_1 \xi^{\pm 2} \eta.
$$
If $c_1 = 0$, then $\eta = 0$ in view of $c_0 \not\in \{-1,1\}$, 
which yields the tubular surfaces $\rho = `const$.
Let us, therefore, assume that $c_1 \ne 0$.
Upon substituting from~\eqref{xi eta}, equation~\eqref{generic:ode}  
becomes a first-order ODE, separable in terms of variables~\eqref{ss}  
and having the elliptic integral
$$
\kappa = \int^\xi \frac{-c_1 t^{\pm2} + c_0^2 - 1}
   {\sqrt{c_1^2 t^{\pm4} - 2 (c_0 + 1) (c_0 + 3) c_1 t^{\pm2}
      + (c_0^2 - 1)^2}}
   \,`d t
$$
as the general solution.
The two cases the `$\pm$' symbol refers to can be converted one into another
by the substitution $c_1 \to (c_0^2 - 1)^2/c_1$.
Therefore, we can safely choose the sign to be `$+$', 
which we do in the sequel. 
Moreover, if $\kappa$ is a solution, then so is $-\kappa$ 
(as a combination of the $\rho \leftrightarrow \sigma$ switch and a scaling 
by factor of $-1$).
This is why we often ignore the sign of~$\kappa$ in what follows.
 
Substituting 
$t \to s/m$, $m = \sqrt{|c_1/(1 - c_0^2)|}$, 
we simplify the integral above to
$$
\numbered\label{Ipm}
\kappa = \frac 1m I_\pm(m\xi,c),
\qquad
I_\pm(\xi,c) = \int^{\xi} \frac{1 \pm s^2}
   {\sqrt{1 + 2 c s^2 + s^4}}
   \,`d s,
$$
where `$\pm$' refers to the signum of $c_1/(1 - c_0^2)$; in particular, is 
unrelated to the `$\pm$' sign in~\eqref{generic:ode}. 
The real parameter $c$ is related to $c_0$ by 
$c = \pm\frac{c_0 + 3}{c_0 - 1}$.

Formula~\eqref{Ipm} describes possible dependences $\rho(\sigma)$ via the
substitution $\kappa = \rho + \sigma$, $\xi = \rho - \sigma$.
Three independent parameters are involved: $m,c$ and the integration constant
(the lower limit of the integral).
Obviously, $m$ plays the role of the scaling parameter. 
The integration constant can be easily identified with the offsetting 
parameter $T$ from~\eqref{offset}. 

Each dependence between $\kappa$ and $\xi$ has a unique representative 
modulo scaling and offsetting, obtainable by fixing the lower limit of the 
integral $I_\pm(\xi,c)$ in~\eqref{Ipm}.
This is straightforward when $c > -1$; we simply redefine
$I_\pm(\xi,c)$ to be
$$
\numbered\label{Ipm_0}
I_\pm(\xi,c) = \int^{\xi}_0 \frac{1 \pm s^2}
   {\sqrt{1 + 2 c s^2 + s^4}}
   \,`d s.
$$
If, however, $c < -1$, then the integrand in~\eqref{Ipm} is real in three 
separate intervals
$(-\infty,-\sqrt{\gamma_+})$, $(-\sqrt{\gamma_-},\sqrt{\gamma_-})$, and 
$(\sqrt{\gamma_+},\infty)$, where
$$
\numbered\label{gamma(c)}
\gamma_\pm = -c \pm \sqrt{c^2 - 1} > 0.
$$ 
We choose the representatives
$-\tilde I_\pm(-\xi,c)$, $I_\pm(\xi,c)$, and $\tilde I_\pm(\xi,c)$,
respectively, 
where $I_\pm(\xi,c)$ is given by~\eqref{Ipm_0} in the interval 
$-\gamma_- \le \xi \le \gamma_-$, while
$$
\numbered\label{tilde Ipm}
\tilde I_\pm(\xi,c) = \int^{\xi}_{\gamma_+} \frac{1 \pm s^2}
   {\sqrt{1 + 2 c s^2 + s^4}}
   \,`d s,
   \quad
   \gamma_+ \le \xi.
$$

\section{Summary of the solutions}
\label{sect:summary}


As demonstrated in the preceding section, each integrable class is determined 
by certain relation between the radii of curvature, which can be subject to 
rescaling 
$\rho \to c_1 \rho$, $\sigma \to c_1 \sigma$, offsetting $\rho \to \rho + c_0$, 
$\sigma \to \sigma + c_0$ and the twist $\rho \leftrightarrow \sigma$.

With the help of Proposition~\ref{prop:uv}, we can find the corresponding 
integrable Gauss equation.
To start with, we investigate the generic class determined by 
formula~\eqref{Ipm}; we fix the scaling for simplicity.

\begin{proposition}
Assuming
$$
\numbered\label{gen}
\rho + \sigma = I_{\pm}(\rho - \sigma,c), \quad  
I_{\pm}(\xi,c) = \int^\xi
  \frac{1 \pm s^2}{\sqrt{1 + 2 c s^2 + s^4}}\,`d s,
$$
the Gauss equation~\eqref{GMC-G} for $\xi = \rho - \sigma$ reads
$$
\numbered\label{Gauss:gen}
R' \xi_{yy} + R'' \xi_y^2 + S' \xi_{xx} + S'' \xi_x^2 + T = 0, 
$$
where
$$
R' = \frac{1 + c \xi^2 + \Delta(\xi,c)}{\xi^2 \Delta(\xi,c)},
\quad
S' = \frac{c \mp 1}2\,\frac{\xi^2}{(1 + c \xi^2 + \Delta(\xi,c)) \Delta(\xi,c)},
\\
\Delta(\xi,c) = \sqrt{1 + 2 c \xi^2 + \xi^4},
\quad 
T = -\frac 1\xi.
$$
The metric coefficients $u,v$ in~\eqref{ff} are
$$
u = \frac{\xi + I_{\pm}(\xi,c)}{2 \xi}
 \sqrt{1 \mp \xi^2 + \Delta(\xi,c)},
\quad
v = \frac{\xi - I_{\pm}(\xi,c)}{2 \xi}
 \sqrt{\frac{1 \mp \xi^2 - \Delta(\xi,c)}{2 c \pm 2}}.
$$
\end{proposition}

\begin{proof}
We parameterize $\rho$ and $\sigma$ by $\xi$, i.e., we resolve~\eqref{gen} as 
$$
\rho = \frac{I_{\pm}(\xi,c) + \xi}2, \quad
\sigma = \frac{I_{\pm}(\xi,c) - \xi}2.
$$
The general form of the Gauss equation, along with the last term 
$T = 1/(\sigma - \rho) = -1/\xi$, follow from Proposition~\ref{prop:uv}.
To find $R',S'$, we compute
$$
(\ln R')' = \frac{R''}{R'}
 = \frac{(\rho - \sigma)\rho'' - 2\rho^{\prime2}}{(\rho - \sigma)\rho'}
 = -\frac2\xi \,
   \frac{c \xi^2 + \xi^4 + \sqrt{1 + 2 c \xi^2 + \xi^4}}{1 + 2 c \xi^2 + \xi^4},
\\
(\ln S')' = \frac{S''}{S'}
 = \frac{(\rho - \sigma)\sigma'' + 2\sigma^{\prime2}}{(\rho - \sigma)\sigma'}
 = -\frac2\xi \,
   \frac{c \xi^2 + \xi^4 - \sqrt{1 + 2 c \xi^2 + \xi^4}}{1 + 2 c \xi^2 + \xi^4}
$$
from~\eqref{RST} under the constraint~\eqref{eq:uv1}. 
These equations need to be integrated once, which is easy;
the integration constants have been chosen to match equations~\eqref{eq:uv1} 
and~\eqref{RST}.
Finally, from~\eqref{RST} one easily computes the coefficients $u,v$ as
$u = \sqrt{{R' \rho^2}/{\rho'}}$, $v = \sqrt{-{S' \sigma^2}/{\sigma'}}$.
\end{proof}

Apart from the generic class we also obtained a number of special solutions, 
listed in Table~\ref{T1} (omitting the tubular surfaces). 
Rows 5b and 6b differ only by translation (offsetting) and
can be identified one with another. 

The first column contains a determining relation (up to a scaling), while
the second harbours the corresponding integrable equation in the compact 
form~\eqref{W:comp}.
Table~\ref{T2} gives the principal radii of curvature $\rho,\sigma$, metric 
coefficients $u,v$, and the variable $z$ (see Table~\ref{T1}) 
in terms of a suitably chosen parameterizing variable $w$. 

\begin{table}
$$
\begin{array}{lll}
   & \text{relation} & \text{integrable equation} \\
\hline
1. & \rho + \sigma = 0 & z_{xx} + z_{yy} + `e^z = 0
\\
\text{2a}. & \rho\sigma = 1 &  z_{xx} + z_{yy} - \sinh z = 0 
\\
\text{2b}. & \rho\sigma = -1 &  z_{xx} - z_{yy} + \sin z = 0 
\\
\text{3a}. & \rho - \sigma = \sinh(\rho + \sigma)  
  & (\tanh z - z)_{xx} + (\coth z - z)_{yy} + `csch 2 z = 0 
\\
\text{3b}. & \rho - \sigma = \sin(\rho + \sigma) 
  & (\tan z - z)_{xx} + (\cot z + z)_{yy} + \csc 2 z = 0
\\
4. & \rho - \sigma = 1 & z_{xx} + (1/z)_{yy} + 2 = 0
\\
\text{5a}. & \rho - \sigma = \tanh \rho 
  & \frac14(\sinh z - z)_{xx} + (\coth \frac12 z)_{yy} + \coth \frac12 z = 0
\\
\text{5b}. & \rho - \sigma = \tan \rho
  & \frac14(\sin z - z)_{xx} + (\cot \frac12 z)_{yy} + \cot \frac12 z = 0
\\ 
\text{6a}. & \rho - \sigma = \coth \rho 
  & \frac14(\sinh z + z)_{xx} - (\tanh \frac12 z)_{yy} + \tanh \frac12 z = 0
\\
\text{6b}. & \rho - \sigma = -\cot \rho   
  & \frac14(\sin z + z)_{xx} + (\tan \frac12 z)_{yy} + \tan \frac12 z = 0
\end{array}
$$
\caption{Special integrable cases and the associated integrable Gauss 
equations}
\label{T1}
\end{table}

\begin{table}
$$
\begin{array}{lccccc}
& \rho & \sigma & u & v & z \\
\hline
1. 
   & w
   & \llap{$-$}w
   & \sqrt{w/2}
   & \sqrt{w/2}
   & \llap{$-$}\ln w
\\
\text{2a}.
   & w 
   & \frac 1w  
   & \frac{w}{\sqrt{w^2 - 1}} 
   & \frac {-1}{\sqrt{w^2 - 1}} 
   & 2 `arctanh w 
\\
\text{2b}.
   & w
   & \llap{$-$}\frac 1w 
   & \frac{w}{\sqrt{w^2 + 1}}
   & \frac 1{\sqrt{w^2 + 1}}
   & 2 `arctan w
\\
\text{3a}.
   & \frac{w + \sinh w}2
   & \frac{w - \sinh w}2
   & \frac{w + \sinh w}{2 \sqrt{\cosh w - 1}}
   & \frac{w - \sinh w}{2 \sqrt{\cosh w + 1}}
   & \frac 12 w 
\\
\text{3b}.
   & \frac{w + \sin w}2
   & \frac{w - \sin w}2
   & \frac{w + \sin w}{2 \sqrt{1 - \cos w}}
   & \frac{w - \sin w}{2 \sqrt{1 + \cos w}}
   & \frac 12 w 
\\
4.
   & w
   & w - 1
   & \frac{w}{`e^{w}}
   & (1 - w) `e^{w}
   & `e^{2w}
\\
\text{5a}.
   & w
   & w - \tanh w
   & \frac{w}{\sinh w}
   & \sinh w - w \cosh w
   & 2 w 
\\
\text{5b}.
   & w
   & w - \tan w
   & \frac{w}{\sin w}
   & \sin w - w \cos w
   & 2 w 
\\
\text{6a}.
   & w
   & w - \coth w
   & \frac{w}{\cosh w}
   & \cosh w - w \sinh w
   & 2 w 
\\
\text{6b}.
   & w
   & w + \cot w
   & \frac{w}{\cos w}
   & \cos w + w \sin w
   & 2 w 
\end{array}
$$
\caption{Special integrable cases. The radii of curvature $\rho, \sigma$,
the metric coefficients $u, v$, and the unknown $z$ of the integrable 
Gauss equation in terms of a variable $w$.}
\label{T2}
\end{table}

Neither of the special cases is new to differential geometry.
Row 1 reflects that, in terms of the curvature line coordinates, minimal 
surfaces correspond to solutions of the Liouville equation~\cite[\S351]{BiaII}.
Similarly, row 2a reproduces the relation between surfaces of negative constant 
Gaussian curvature and solutions of the elliptic sinh-Gordon equation.
Row 2b does the same for the hyperbolic sine-Gordon equation and surfaces
of positive constant Gaussian curvature 
(or constant mean curvature, by the theorem of Bonnet on parallel surfaces).
Nowadays, surfaces of constant mean or Gaussian curvature are undoubtedly the 
best understood classes of surfaces integrable in the sense of soliton theory 
(see, e.g.,~\cite{Bob1,Bob4,D-N,Hel,Me-S,P-St} and references therein).

It may come as a surprise that the other cases are classical as well.
Introduced by Weingarten~\cite[\S4]{Wein2} 
(`eine neue Fl\"achenklasse'), 
surfaces satisfying the relation $\rho - \sigma = \sin(\rho + \sigma)$ 
(row 4b) are covered in 
Darboux~\cite[\S\S745, 746, 766, 769, 770]{DarIII} 
(`une classe nouvelle de surfaces d\'ecouverte par M.~Weingarten') 
and Bianchi~\cite[\S135]{BiaI}, \cite[\S245]{BiaII}. 
Darboux~\cite[\S746]{DarIII} gave a general solution of an equation equivalent
to our
$(\tan z - z)_{xx} + (\cot z + z)_{yy} + \csc 2 z = 0$. 
He also provided a remarkable geometric construction in~\cite[\S770]{DarIII}, 
further developed by Bianchi~\cite[\S245]{BiaII}. 
In a nutshell: the middle evolutes are translation surfaces generated by curves
of opposite constant nonzero torsion; conversely the Weingarten surfaces are 
orthogonal to the osculation planes of the generating curves. 
Bianchi's research extends to the complementary relation
$\rho - \sigma = \sinh(\rho + \sigma)$ (row 3a) as well~\cite[\S246]{BiaII}.
The remaining rows (from 4 to 6b) correspond to involutes of surfaces of 
constant Gaussian curvature studied by Beltrami~\cite[Ch.~9, \S20]{Bel}. 
Row~4 (surfaces of constant astigmatism) has been addressed in Part~I; 
we have nothing to add except the Beltrami's work as the earliest reference we 
know of.

Table~\ref{Tlim} demonstrates how the cases expressible in terms of elementary 
functions arise as limits of the generic integral~\eqref{Ipm} for $c$ approaching 
$\pm 1$ or $\pm\infty$ along a suitable curve in the $(c,m)$ space. 
The tubular surfaces $\sigma = `const$, which are omitted, correspond to 
$\kappa = I_+(\xi,1) = \xi + `const$. 

\begin{table}
$$
\begin{array}{lll}
   & \text{relation} & \text{limit} \\
\hline
1. & \kappa = 0 & I_\pm (\xi, \infty) 
\\
\text{2a}. & \kappa^2 = \xi^2 + 4 &  \lim_{m = \infty} I_\pm (m\xi, 2 m^2)/m
\\
\text{2b}. & \kappa^2 = \xi^2 - 4 &  \lim_{m = \infty} I_\pm (m\xi, -2 m^2)/m 
\\
\text{3a}. & \kappa = `arcsinh \xi  
  & \lim_{m = 0} I_\pm (m\xi, 1/{2 m^2})/m
\\
\text{3b}. & \kappa = `arcsin \xi 
  & \lim_{m = 0} I_\pm (m\xi, -1/{2 m^2})/m
\\
4. & \xi = 1 & \lim_{m = \infty} \tilde I_\pm (m\xi, -m^2/2)/m  
\\
\text{5a}. & \kappa = -\xi + 2`arctanh \xi
  & I_+(\xi,-1),\ |\xi| \lt 1
\\
\text{5b}. & \kappa = -\xi + 2`arctan \xi
  & I_-(\xi,1)
\\ 
\text{6a}. & \kappa = -\xi + 2`arccoth \xi
  & I_+(\xi,-1),\ |\xi| \gt 1
\\
\text{6b}. & \kappa = -\xi - 2`arccot \xi  
  & I_-(\xi,1)
\end{array}
$$
\caption{Special integrable cases as limits of $I_{\pm}(\xi,c)$}
\label{Tlim}
\end{table}

\section{Curvature diagrams}

To exemplify the wealth of classes of integrable surfaces, we plot 
the representative solutions of the governing equation~\eqref{ODE} 
in Figures~1 and~2. 
We call them curvature diagrams, even though the radii of curvature 
$\rho,\sigma$, rather than the curvatures $1/\rho,1/\sigma$, are 
plotted, contrary to the customary practice~\cite[Ch.~5]{H-C}.
The benefit is that diagrams can be not only scaled arbitrarily,
but also freely translated along the dashed line $\rho = \sigma$; 
the translation corresponds to offsetting. 
For clarity, we adjusted the offsetting so that the diagrams are 
symmetric about the origin, i.e.,
$\rho(\sigma) = -\rho(-\sigma)$.

The diagrams contain plots of functions 
$\mathcal I_{`A}(\xi,k)$, 
$\mathcal{I}_{`B}(\xi,k)$, 
$\mathcal{I}_{`C\pm}(\xi,k)$,
and $k\tilde{\mathcal I}_{`A}(\xi/k,k)$.
All special cases are explicitly included as limits, except the 
surfaces of constant positive curvature (row~2a). 
These could be obtained as the limit of
$k \mathcal{I}_{`B}(\xi/k, k)$ as $k$ approaches zero.

\begin{figure} 
\begin{center}
  \includegraphics[width=2\unitlength]{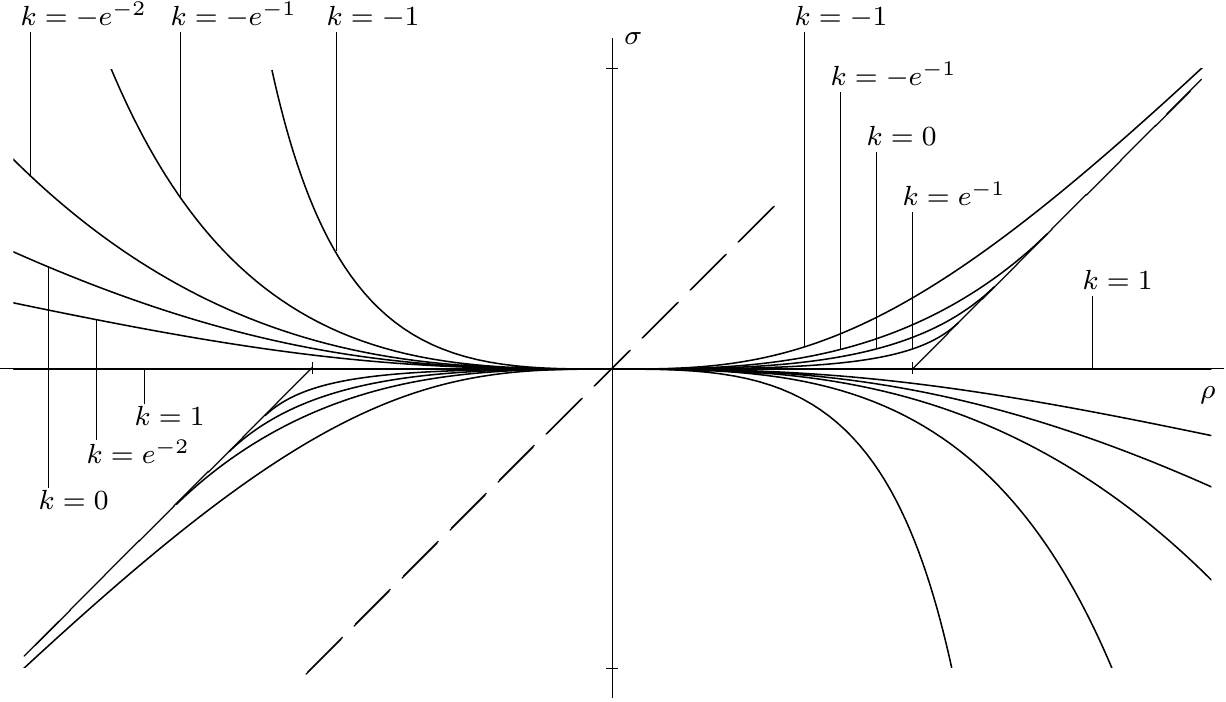}
  \caption{Curvature diagrams $\kappa = \mathcal{I}_{`B}(\xi, k)$ 
	(the left-hand legend) and 
    $\kappa = \mathcal{I}_{`A}(\xi,k)$, $|\xi| < 1$ 
    (the right-hand legend), 
    where $\kappa = \rho+\sigma$, $\xi = \rho - \sigma$. 
	More can be obtained by rescaling and translating along the dashed 
	line $\rho = \sigma$, the axis $\kappa$.
	Here
    $\mathcal{I}_{`A}(\xi,-1) = -\xi + 2 `arctan \xi$ (row~5b), 
    $\mathcal{I}_{`A}(\xi,0) = `arcsin \xi$ (row~3b),
    $\mathcal{I}_{`A}(\xi,1) = \xi$;
    $\mathcal{I}_{`B}(\xi,-1) = -\xi + 2 `arctanh \xi$ (row~5a), 
    $\mathcal{I}_{`B}(\xi,0) = `arcsinh \xi$ (row~3a),
    $\mathcal{I}_{`B}(\xi,1) = \xi$.
    Graphs of $\kappa = \mathcal{I}_{`A}(\xi,k)$ 
    end on the solid lines $|\xi| = 1$.}
\end{center}
\end{figure} 

\begin{figure} 
\begin{center}
\begin{picture}(1,1)(0,0) 
\put(0.5,0.5){\makebox(0,0){\includegraphics[width=\unitlength]{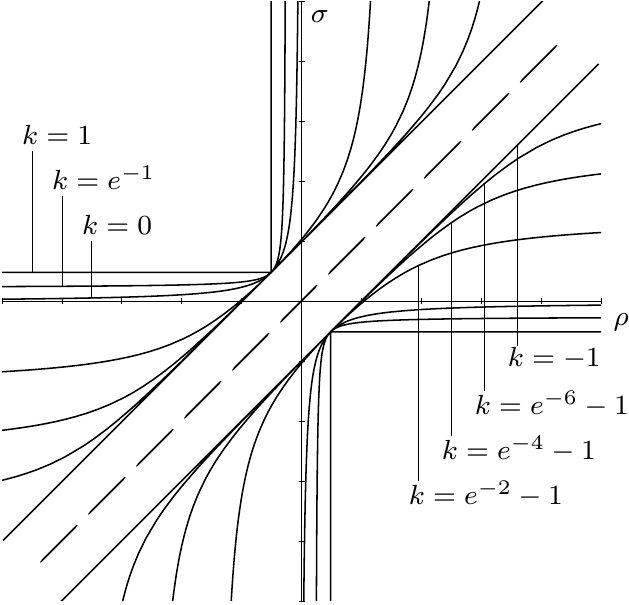}}} 
\put(0,0){\makebox(0,0)[lb]{\scriptsize(a)}} 
\end{picture}
\quad
\begin{picture}(1,1)(0,0) 
\put(0.5,0.5){\makebox(0,0){\includegraphics[width=\unitlength]{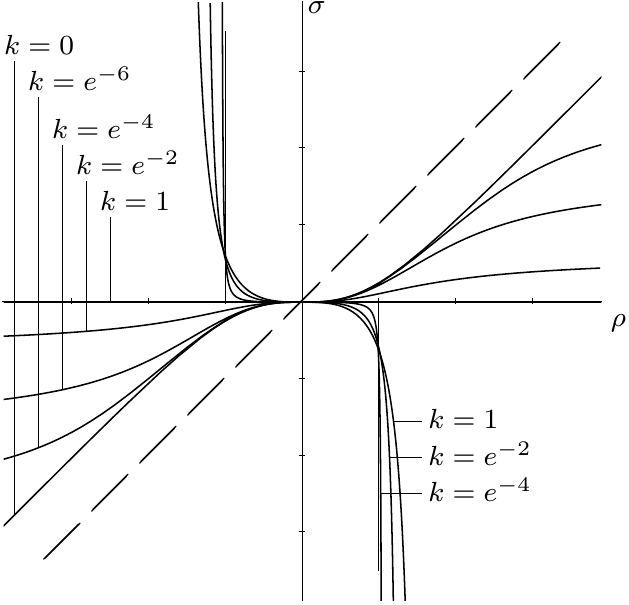}}} 
\put(0,0){\makebox(0,0)[lb]{\scriptsize(b)}} 
\end{picture}
\end{center}
\caption{Curvature diagrams  
(a)  $\kappa =  k \tilde{\mathcal{I}}_{`A}(\xi/k, k)$,  
     $|\xi|> 1/|k|$;  
(b)  $\kappa = \mathcal{I}_{C+}(\xi,k)$ 
     (the top left-hand legend)   
     and $\kappa = -\mathcal{I}_{C-}(\xi,k)$ 
     (the bottom right-hand legend), 
     where $\kappa = \rho+\sigma$, $\xi = \rho - \sigma$.
     More can be obtained by rescaling and translating along the dashed 
	 line $\rho = \sigma$, the axis $\kappa$.
	 \\
	 In (a), the line $k = 1$ corresponds to tubular surfaces, 
	 $k = 0$ to surfaces of negative constant curvature (row~2b),
	 and $k = -1$ to the constant astigmatism surfaces (row~4).
	 In (b), $\mathcal{I}_{C+}(\xi,0) = -\xi + 2 `arctan \xi$ (row~5b) 
	 $\mathcal{I}_{C-}(\xi,1) = \xi - 2`arctan((\xi - 1)/(\xi + 1))$ 
	 (row~5a after reparameterization).
	 }
\end{figure} 

The plots have been calculated using the Legendre 
normal form~\cite{Erd,P-So} of the elliptic integrals~\eqref{Ipm_0} 
and~\eqref{tilde Ipm}, which could be of independent interest.
As is well known, the Legendre normal form depends on the 
configuration of roots of 
the quartic polynomial $\Pi = s^4 + 2 c s^2 + 1$.

A) If $c < -1$, then $\Pi = (s^2 - \gamma_+)(s^2 - \gamma_-)$ has four 
real roots $\sqrt{\gamma_\pm}$ and $-\sqrt{\gamma_\pm}$ given by 
formula~\eqref{gamma(c)}.
By using the substitution $s = \sqrt{k} r$, where $k = \gamma_-$, 
we easily obtain the Legendre normal form 
$$
\frac1{\sqrt{k}} I_\pm(\xi \sqrt{k},-\frac{k^2 + 1}{2 k})
  =  \int_0^{\xi}
    \frac{1 \pm k r^2}{\sqrt{(1 - r^2)(1 - k^2 r^2)}}\,`d r,
\quad 0 < k < 1.
$$
On the right-hand side, we can remove the $\pm$ sign from the 
numerator by allowing $k$ to range between $-1$ and~$1$.
For $-1 \le \xi \le 1$, $-1 < k < 1$,
we have a unified representative given by 
$\kappa = \mathcal I_{`A}(\xi,k)$, 
where
$$
\mathcal I_{`A}(\xi,k)
 = \int_0^{\xi}
    \frac{1 - k r^2}{\sqrt{(1 - r^2)(1 - k^2 r^2)}}\,`d r
 = \frac{1}{k} E(\xi; k) + \frac{k - 1}{k} F(\xi; k)
$$
in terms of the Legendre elliptic integrals $E,F$. 

For real $\xi$ such  that $|\xi| > 1$,
the function $\mathcal I_{`A}(\xi,k)$ is complex valued.
Yet we obtain a real function for $1/|k| \le \xi$ by choosing
the lower limit of the integral to be $1/k$, $-1 < k < 1$.
Thus, 
$$
\tilde{\mathcal I}_{`A}(\xi,k)
 = \begin{cases}
    \int_{1/|k|}^{\xi}
    \frac{1 - k r^2}{\sqrt{(1 - r^2)(1 - k^2 r^2)}}\,`d r
    = \mathcal I_{`A}(\xi,k) - \mathcal I_{`A}(\frac1{|k|},k), & \xi > \frac1{|k|},
    \\
    -\tilde{\mathcal I}_{`A}(-\xi,k), & \xi < -\frac1{|k|}.  
\end{cases}
$$


B) Similarly, when $c > 1$, then $\gamma_\pm \lt 0$, the roots 
$\sqrt{\gamma_\pm}$, $-\sqrt{\gamma_\pm}$ of $\Pi$ are purely imaginary,
and
$$
\frac1{\sqrt{k}} I_\pm(\xi \sqrt{k}, \frac{k^2 + 1}{2 k})
  = \int_0^{\xi}
    \frac{1 \pm k r^2}{\sqrt{(1 + r^2)(1 + k^2 r^2)}}\,`d r,
\quad 0 < k < 1.
$$
The two representatives can be unified into $\kappa = \mathcal I_{`B}(\xi,k)$,
where
$$
\mathcal I_{`B}(\xi,k) = \int_0^{\xi}
    \frac{1 - k r^2}{\sqrt{(1 + r^2)(1 + k^2 r^2)}}\,`d r
 = \frac{1}{k`i} E(\xi`i; k) + \frac{k - 1}{k`i} F(\xi`i; k)
$$
for $-1 < k < 1$.

C) When $-1 < c < 1$ (four distinct complex roots), we substituted
$$
s = \frac{1 + \sqrt{k} r}{1 - \sqrt{k} r},
\quad 0 < k < 1,
$$
to obtain two more representatives $\kappa = \mathcal I_{`C+}$ and 
$\kappa = \mathcal I_{`C-}$, 
where
$$
\mathcal I_{`C\pm} = \begin{cases}
   J_{`C\pm}(\xi,k) - J_{`C\pm}(0,k), & \xi \ge 0, \\
   -\mathcal I_{`C\pm}(-\xi,k), & \xi \lt 0,
   \end{cases} 
\\
J_{`C\pm}(\xi,k) = \frac{\sqrt{1 + 2 c \xi^2 + \xi^4}}{1 + \xi}
 + \frac{2}{(k + 1)`i} E(\frac{\xi - 1}{\xi + 1}\frac{`i}{\sqrt k}, k)
 + \frac{\epsilon_\pm}{`i} F(\frac{\xi - 1}{\xi + 1}\frac{`i}{\sqrt k}, k),
\\
\epsilon_\pm = \frac{(1 \pm 1)k - 3 \pm 1}{2}
 = \begin{cases} k - 1, \\ -2, \end{cases}
\\
c = -\frac{k^2 - 6 k + 1}{(k+1)^2}.
$$

\section{Normal congruences and their focal surfaces}
\label{sect:focal}


The fact that the governing equation~\eqref{ODE} has the offsetting 
symmetry~\eqref{offset} is not a pure coincidence.
Being invertible, the offsetting transformation 
$\mathbf r \mapsto \mathbf r + T \mathbf n$ preserves integrability 
in every reasonable sense of the word. 
Surfaces related by the offsetting transformation 
are said to be parallel and either all are integrable or none is.
However, parallel surfaces can be alternatively described as normal 
surfaces to the same line congruence. 
Consequently, integrability is a property of this congruence and,
therefore, must have an expression in terms of congruence 
invariants. 

Normal congruences of Weingarten surfaces, also known as 
$W$-congruences, are rather special with regard to properties of 
their focal surfaces. 
It is therefore natural to look for characterization of the former 
in terms of the latter.
Naturally, we expect the focal surfaces of integrable $W$-congruences 
to be integrable as well.

Recall that a generic surface has two focal surfaces 
(often considered as two sheets of a single surface), 
$$
\mathbf r^{(1)} = \mathbf r + \sigma \mathbf n, \qquad
\mathbf r^{(2)} = \mathbf r + \rho \mathbf n.
$$
each of which is formed by the evolutes of one family of 
the curvature lines. 
Focal surfaces can degenerate into a line or even a point.
In the case of a Weingarten surface $\mathbf r$ with fundamental 
forms~\eqref{ff}, one of the focal surfaces degenerates into a line 
if $\sigma_y = \sigma' w_y = 0$ or $\rho_x = \rho' w_x = 0$; 
both degenerate into a point if the surface is a sphere 
(already excluded from consideration); 
otherwise they are regular surfaces.
Therefore, we assume $\rho' \sigma' \ne 0$ in what follows.

To compute the respective first and second fundamental forms 
$`I^{(i)}$ and $`II^{(i)}$, $i = 1,2$, we proceed as follows.
In view of the Gauss--Mainardi--Codazzi equations~\eqref{GMC-G} 
and~\eqref{GMC-MC}, the Gauss--Weingarten~\eqref{GW} equations can be 
written as
$$
\numbered \label{GW*}
\mathbf r_{xx} = \frac{u_x}{u} \mathbf r_x
  + \frac{\sigma \rho' u^2 w_y}{\rho (\rho - \sigma) v^2} \mathbf r_y
  + \frac{u^2}{\rho} \mathbf n, 
\qquad
\mathbf n_x = -\frac1\rho \mathbf r_x, \\
\mathbf r_{xy} = \frac{\sigma \rho' w_y}{\rho(\sigma - \rho)} \mathbf  
r_x
  + \frac{\rho \sigma' w_x}{\sigma(\rho - \sigma)} \mathbf r_y, \\
\mathbf r_{yy} = \frac{\rho \sigma' v^2 w_x}{\sigma (\sigma - \rho)  
u^2} \mathbf r_x
  + \frac{v_y}{v} \mathbf r_y
  + \frac{v^2}{\sigma} \mathbf n, \qquad
\mathbf n_y = -\frac1\sigma \mathbf r_y.
$$ 
One easily finds
$$
\mathbf r^{(1)}_x = \frac{\rho - \sigma}\rho \mathbf r_x
 + \sigma' w_x \mathbf n,
\quad
\mathbf r^{(1)}_y = \sigma' w_y \mathbf n,
\quad
\mathbf n^{(1)} = \frac{\mathbf r_y}{v},
\\
\mathbf r^{(2)}_x = \rho' w_x \mathbf n,
\quad
\mathbf r^{(2)}_y = \frac{\sigma - \rho}\sigma \mathbf r_y
 + \rho' w_y \mathbf n,
\quad
\mathbf n^{(2)} = \frac{\mathbf r_x}{u}.
$$
Using the equations~\eqref{GW*} and~\eqref{ff}, we get
$$
\numbered\label{evol:ff1}
`I^{(1)} = \frac{(\rho - \sigma)^2 u^2}{\rho^2}`d x^2 +\, `d\sigma^2, \qquad
`I^{(2)} = d\rho^2 + \frac{(\rho - \sigma)^2 v^2}{\sigma^2}`d y^2, 
$$
where $`d\rho = \rho'\,`d w = \rho'(w_x\,`d x + w_y\,`d y)$, 
$`d\sigma = \sigma'\,`d w = \sigma'(w_x\,`d x + w_y\,`d y)$. 

With $u,v$ determined from Proposition~\ref{prop:uv}, we can write
$$
`I^{(1)} = (f^{(1)}(\sigma)\,`d x)^2 +\, `d\sigma^2,  \qquad
`I^{(2)} = (f^{(2)}(\rho)\,`d y)^2 +\, `d\rho^2.
$$ 
Hence, all focal surfaces 
$\mathbf r^{(i)}$ corresponding to a given dependence $\rho(\sigma)$ are 
isometric. Moreover, the first fundamental 
forms~\eqref{evol:ff1} are typical of surfaces of revolution.
These are among the classical results by Weingarten~\cite{Wein2}.

Omitting details, we further compute the second fundamental forms
$$
\numbered\label{evol:ff2}
`II^{(1)} = \frac{\sigma w_y}{v} (\frac{\rho' u^2}{\rho^2}`d x^2
 - \frac{\sigma' v^2}{\sigma^2}`d y^2), \quad
`II^{(2)} = -\frac{\rho w_x}{u} (\frac{\rho' u^2}{\rho^2}\,`d x^2
 - \frac{\sigma' v^2}{\sigma^2}`d y^2)
$$
and note that they are conformally related, which is another way to 
express Ribaucour's classical result~\cite{Ri} that asymptotic 
coordinates on $\mathbf r^{(1)}$ and $\mathbf r^{(2)}$ correspond.
The Gaussian curvatures are
$$
\numbered\label{evol:GC}
K^{(1)} = \frac{\det `II^{(1)}}{\det `I^{(1)}}
 = -\frac{\rho'}{(\rho - \sigma)^2 \sigma'},
\qquad
K^{(2)} = \frac{\det `II^{(2)}}{\det `I^{(2)}}
 = -\frac{\sigma'}{(\rho - \sigma)^2 \rho'}.
$$
Consequently, the focal surfaces have one and the same sign 
of the Gaussian curvature, which we denote as $\epsilon$.
We have $\epsilon = -1$ (both focal surfaces are hyperbolic) 
if and only if $`d\rho/`d\sigma = \rho'/\sigma' > 0$ 
(if $\rho$ increases as $\sigma$ increases), and $+1$ if
$`d\rho/`d\sigma < 0$. 
The relation 
$$
\numbered\label{Hal}
K^{(1)} K^{(2)} = \frac 1{(\rho - \sigma)^4}
$$
away of umbilic points is known as the Halphen theorem 
(see~\cite[\S129]{BiaI}).

As we have already explained, to every particular relation $\rho(\sigma)$ of 
curvatures there corresponds an isometry class of focal surfaces, which 
contains a unique rotational representative (which is the way the classes 
have been characterized in the classical literature). 
However, we believe that a description in terms of metric invariants is more 
appropriate. 
It is convenient to choose
$$
\kappa^{(i)} = \frac1{\sqrt{\epsilon K^{(i)}}},
$$
where $\epsilon K^{(i)} = |K^{(i)}|$ is the absolute value of the Gaussian 
curvature of the $i$th focal surface.

Further, let $\gamma^{(i)}$ be defined by
$$
\numbered\label{gamma12}
\gamma^{(1)} = 
  \frac{(\rho - \sigma) (\rho'' \sigma' - \sigma'' \rho')
   - 2 \rho' \sigma' (\rho' - \sigma')}{2\,(-\epsilon \rho' \sigma')^{3/2}},
 \\
\gamma^{(2)} = 
   \frac{(\rho - \sigma) (\rho'' \sigma' - \sigma'' \rho')
   + 2 \rho' \sigma' (\rho' - \sigma')}{2\,(-\epsilon \rho' \sigma')^{3/2}}.
$$
One can directly check that $|\gamma^{(i)}|$
equals the norm of the gradient of $\kappa^{(i)}$ with respect to~$`I^{(i)}$,
$$
|\gamma^{(i)}|
 = \left\|`grad^{(i)}\kappa^{(i)}\right\|^{(i)}
 = \sqrt{I^{(i)} (`grad^{(i)}\kappa^{(i)}, `grad^{(i)}\kappa^{(i)})}.
$$ 
Hence, $\gamma^{(i)}$ is a metric invariant of the respective focal surface.
It is sometimes more convenient to use invariants
$$
\numbered\label{G12}
G^{(1)} = 
   \frac{[(\rho - \sigma) (\rho'' \sigma' - \sigma'' \rho')
   - 2 \rho' \sigma' (\rho' - \sigma')]^2}{16\,(\rho' \sigma')^3},
 \\
G^{(2)} = 
   \frac{[(\rho - \sigma) (\rho'' \sigma' - \sigma'' \rho')
   + 2 \rho' \sigma' (\rho' - \sigma')]^2}{16\,(\rho' \sigma')^3},
$$
satisfying
$$
\gamma^{(i)2} = -4\epsilon G^{(i)},
\quad
-16\,G^{(i)} K^{(i)3} = I^{(i)} (`grad^{(i)} K^{(i)}, `grad^{(i)} K^{(i)}).
$$

Clearly, both $\kappa^{(i)}$ and $G^{(i)}$ are functions of $w$. 
Consequently, $G^{(i)}$ can be considered as a function of $\kappa^{(i)}$ 
unless $\kappa^{(i)}$ is a constant. 
Our nearest aim is to establish the dependence between $\kappa^{(i)}$ and 
$G^{(i)}$ in terms of the dependence between $\rho$ and $\sigma$.

\begin{proposition}
\label{prop:focal}
Let the principal radii of curvature $\rho,\sigma$ of an integrable
surface satisfy the generic relation~\eqref{Ipm}. 
Then the metric invariants $G^{(i)}$ and $\kappa^{(i)}$ satisfy the
relations
$$
\numbered\label{Gkappa}
G^{(1)} = (-1 \pm \sqrt{\frac 2{c \mp 1}} \frac{\kappa^{(1)}}m)
 (1 + \sqrt{\frac 2{c \mp 1}} \frac m{\kappa^{(1)}}),
\\
G^{(2)} = (1 \pm \sqrt{\frac 2{c \mp 1}} \frac{\kappa^{(2)}}m)
 (-1 + \sqrt{\frac 2{c \mp 1}} \frac1 m{\kappa^{(2)}}).
$$
Furthermore,
$$
G^{(1)} G^{(2)} = (\frac{c \pm 1}{c \mp 1})^2
$$
is constant (hence, so is the product $\gamma^{(1)} \gamma^{(2)}$).

Table~\ref{T4} lists the product $G^{(1)} G^{(2)}$ and the algebraic
relations between $G^{(i)}$ and $\kappa^{(i)}$ in the special cases.
 \end{proposition}

\begin{proof}
For simplicity, we start assuming a fixed scaling, i.e., we depart
from formula~\eqref{gen}.
We routinely compute
$$
K^{(1)} = 
  \frac{(1 \pm w^2 + \sqrt{1 + 2c w^2 + w^4})^2}
       {2 (c \mp 1) w^4},
\quad
K^{(2)} = 
  \frac{(1 \pm w^2 - \sqrt{1 + 2c w^2 + w^4})^2}
       {2 (c \mp 1) w^4}.
$$
Consequently, $\epsilon = `sgn(c \mp 1))$, and
$$
\kappa^{(1)} =  
  \frac{1 \pm w^2 - \sqrt{1 + 2c w^2 + w^4}}
       {\sqrt{2|c \mp 1|}},
\quad
\kappa^{(2)} =  
  \frac{1 \pm w^2 + \sqrt{1 + 2c w^2 + w^4}}
       {\sqrt{2|c \mp 1|}}.
$$
Furthermore,
$$
G^{(1)} =  
  -\frac{(1 \mp w^2 + \sqrt{1 + 2 c w^2 + w^4})^2}
       {2 (c \mp 1)w^2},
\quad
G^{(2)} =  
  -\frac{(1 \mp w^2 - \sqrt{1 + 2 c w^2 + w^4})^2}
       {2 (c \mp 1)w^2}.
$$
Under the scaling by factor of $m$, the metric invariants 
$K^{(i)}$ and $\kappa^{(i)}$ become $K^{(i)}/m^2$ and 
$m \kappa^{(i)}$, respectively,
while $G^{(i)}$ remains invariant. 
Formulas~\eqref{Gkappa} are then easily checked.
Moreover, all three metric invariants are invariant under the 
offsetting~\eqref{offset}.

Formulas for $G^{(i)}$ and $\kappa^{(i)}$ in the special cases 
are given in Table~\ref{T3} along with the sign $\epsilon$ of 
the Gaussian curvatures.
\end{proof}

\begin{table}
$$
\begin{array}{lrcccc}
& \epsilon & \kappa^{(1)} & \kappa^{(2)} & G^{(1)} & G^{(2)} \\
\hline
1. 
   & 1
   & \hm 2\,|w|
   & \hm 2\,|w|
   & -1
   & -1
\\
\text{2a}. 
   & 1
   & \hm \left|\frac 1{w^2} - 1\right|  
   & \hm |w^2 - 1| 
   & -\frac 1{w^2} 
   & -w^2
\\
\text{2b}. 
   & -1
   & \hm \frac 1{w^2} + 1 
   & \hm w^2 + 1 
   & \hm\frac 1{w^2} 
   & \hm w^2
\\
\text{3a}. 
   & 1
   &  -1 + \cosh w
   & \hm 1 + \cosh w
   & \hm\frac{1 + \cosh w}{1 - \cosh w}
   & \hm\frac{1 - \cosh w}{1 + \cosh w}
\\
\text{3b}. 
   & -1
   & \hm 1 - \cos w
   & \hm 1 + \cos w
   & \hm\frac{1 + \cos w}{1 - \cos w}
   & \hm\frac{1 - \cos w}{1 + \cos w}
\\
4. 
   & -1
   & \hm 1
   & \hm 1
   & \hm 0
   & \hm 0
\\
\text{5a}. 
   & -1
   & \hm \tanh^2 w
   & \hm 1
   & \hm\frac 1{\sinh^2 w \cosh^2 w}
   & \hm 0
\\
\text{5b}. 
   & 1
   & \hm \tan^2 w
   & \hm 1
   & -\frac 1{\sin^2 w \cos^2 w}
   & \hm 0
\\
\text{6a}. 
   & -1
   & \hm \coth^2 w
   & \hm 1
   & \hm\frac 1{\sinh^2 w \cosh^2 w}
   & \hm 0
\\
\text{6b}. 
   & 1
   & \hm \cot^2 w
   & \hm 1
   & -\frac 1{\sin^2 w \cos^2 w}
   & \hm 0
\end{array}
$$
\caption{Special integrable cases. 
Metric invariants of focal surfaces in terms of $w$.}
\label{T3}
\end{table}

\begin{table}
$$
\begin{array}{lrcccc}
& \epsilon & G^{(1)} G^{(2)} & G^{(1)}(\kappa^{(1)}) & 
 G^{(2)}(\kappa^{(2)}) \\
\hline
1. 
   & 1
   & -1 
   & -1
   & -1
\\
\text{2a}. 
   & 1
   & \hm 1
   & -1 \pm \kappa^{(1)}
   & -1 \pm \kappa^{(2)}
\\
\text{2b}. 
   & -1
   & \hm 1 
   & -1 + \kappa^{(1)}
   & -1 + \kappa^{(2)}
\\
\text{3a}. 
   & 1
   & -1 
   & -1 - \frac2{\kappa^{(1)}}
   & -1 - \frac2{\kappa^{(2)}}
\\
\text{3b}. 
   & -1
   & \hm 1 
   & -1 + \frac2{\kappa^{(1)}} 
   & -1 + \frac2{\kappa^{(2)}}
\\
4. 
   & -1
   & \hm 0
   & \hm 0
   & \hm 0
\\
\text{5a}. 
   & -1
   & \hm 0
   & \hm (\sqrt{\kappa^{(1)}} - \frac1{\sqrt{\kappa^{(1)}}})^2
   & \hm 0
\\
\text{5b}. 
   & 1
   & \hm 0
   & -(\sqrt{\kappa^{(1)}} + \frac1{\sqrt{\kappa^{(1)}}})^2
   & \hm 0
\\
\text{6a}. 
   & -1
   & \hm 0
   & \hm (\sqrt{\kappa^{(1)}} - \frac1{\sqrt{\kappa^{(1)}}})^2
   & \hm 0
\\
\text{6b}. 
   & 1
   & \hm 0
   & -(\sqrt{\kappa^{(1)}} + \frac1{\sqrt{\kappa^{(1)}}})^2
   & \hm 0
\end{array}
$$
\caption{Special integrable cases. 
Relations between metric invariants of focal surfaces.}
\label{T4}
\end{table}


Summarizing, focal surfaces of integrable Weingarten surfaces belong 
to the isometry classes specified in Proposition~\ref{prop:focal}.

A natural question is whether the condition $G^{(1)} G^{(2)} = `const$
or, equivalently, $\gamma^{(1)} \gamma^{(2)} = `const$, is not only
necessary, but also sufficient for the condition~\eqref{ODE} to hold.

\begin{proposition}
\label{prop:GG}
Under the condition $\gamma^{(1)} + \gamma^{(2)} \ne 0$, a surface
satisfies the governing equation~\eqref{ODE} 
if and only if the product 
$$
\numbered\label{GG}
\gamma^{(1)} \gamma^{(2)}
 = \pm\left\|`grad^{(1)}\kappa^{(1)}\right\|^{(1)}\,
   \left\|`grad^{(2)}\kappa^{(2)}\right\|^{(2)}
$$
is constant.
\end{proposition}

\begin{proof}
Assuming the $\rho(\sigma)$ dependence, 
$\gamma^{(1)} + \gamma^{(2)}$ simplifies to $(\rho - \sigma) \rho''/\sqrt{|\rho'|^3}$ 
and the product in question to 
$$
\gamma^{(1)} \gamma^{(2)} = \frac{(\rho' - 1)^2}{\epsilon\rho'}
 - \frac{(\rho - \sigma)^2 \rho^{\prime\prime 2}}{4\,\epsilon\rho^{\prime3}}.
$$
Factorizing the $\sigma$-derivative of this expression as 
$$
\pm\frac{(\rho - \sigma)^2}{2\,\epsilon\rho^{\prime3}}
(\rho''' - \frac{3}{2 \rho'} \rho''{}^2
  + \frac{\rho' - 1}{\rho - \sigma} \rho''
  - 2 \frac{(\rho' - 1) \rho' (\rho' + 1)}{(\rho - \sigma)^2})
\rho'' 
$$ 
and comparing to the governing equation~\eqref{ODE} proves the proposition.
\end{proof}

It follows from the proof that condition~\eqref{GG} also 
holds when $\rho'' = 0$, i.e., if there is a linear relation between the 
radii of curvature. 
As of now, there seems to be no indication towards integrability of the 
latter class (except when $\rho \pm \sigma = `const$, which 
satisfies~\eqref{ODE} as well).

\section{Conclusions and future work}

In this work we singled out a class of Weingarten surfaces on the 
basis of its solitonic integrability.
Although special cases were not unknown to nineteenth century geometers, 
the overall result appears to be new. 
We also characterized integrability in terms of metric 
invariants of the focal surfaces.

For time reasons, many questions had to be left for further research. 
We do not know the B\"acklund transformation, recursion operator, 
bi-Hamiltonian structure and other attributes of integrability. 
We did not provide any solutions to the Gauss equation~\eqref{Gauss:gen}.
We do not know what is the true geometric meaning of the spectral 
parameter.
Even the task of computing third order symmetries of the Gauss
equation proved to be too complex.

We have seen in Part~I that integrability of surfaces of constant 
astigmatism is attributable to the fact that their focal surfaces are 
pseudospherical. 
In the general case, the existence of an integrability-preserving 
relation to previously known integrable surfaces is an open problem.

Our nearest goals include exploring the induced Bianchi type 
transformation between surfaces satisfying relations~\eqref{Gkappa}
as well as investigating the extended symmetries of the class in the
sense of Cie\'sli\'nski~\cite{C-nls,C-G-S}.

\ack

We are indebted to J.~Cie\'sli\'nski, E.~Ferapontov, A.~Sergyeyev, 
and S.~Verpoort for advice and encouragement.
The first-named author was supported by GA\v{C}R under project
201/07/P224;
the second-named author by M\v{S}MT under project MSM~4781305904
``Topologick\'e a analytick\'e metody v teorii
dynamick\'ych syst\'em\accent23u a matematick\'e fyzice.''
Thanks are also due to CESNET for granting access to the MetaCentrum
computing facilities.

\end{document}